\documentclass[amsmath,superscriptaddress,aps,pra,twocolumn]{revtex4-2}

\usepackage[T1]{fontenc}
\usepackage{lmodern}
\usepackage{microtype}
\usepackage{graphicx}
\usepackage{bm}
\usepackage{siunitx}
\usepackage{amssymb}
\usepackage{amsmath}
\usepackage{microtype}

\usepackage{chemformula}
\usepackage{siunitx}
\usepackage{gensymb}

\usepackage[colorlinks=true,linkcolor=blue,citecolor=black,urlcolor=black]{hyperref}
\graphicspath{{images/}}

\begin{document}

\title{Trifolium nanocavity metasurfaces on single-crystal Au(111) for depth-tunable optical-variable reflection}

\author{Amos Sospeter Kiyumbi}
\email{amos.kiyumbi@udsm.ac.tz}
\affiliation{Department of Physics, Mathematics and Informatics, Dar es Salaam University College of Education, P.O. Box 2329, Dar es Salaam, Tanzania}
\affiliation{Institute of Experimental Physics, Ulm University, Albert-Einstein-Allee 11, 89069 Ulm, Germany}

\begin{abstract}
Symmetry-broken plasmonic nanocavities provide a simple route to engineer reflective optical response in continuous-metal metasurfaces. Here, we report an experimental study of trifolium-shaped nanocavity arrays milled into single-crystal Au(111) microplates and characterized by white-light reflection spectroscopy in the visible--near-infrared. The structured Au surfaces exhibit broad but well-defined reflection bands and pronounced low-reflectance regions that differ strongly from flat gold. We show that the optical response is highly sensitive to groove depth: increasing the cavity depth from \SI{300}{nm} to \SI{350}{nm} induces a clear redshift ($\sim$ \SI{63}{nm}) of the dominant long-wavelength minimum band ($\lambda = 700-800$ nm) and reshapes the intermediate spectral profile. In addition, the trifolium geometry shows a measurable azimuth-dependent response under sample rotation, unlike the azimuthally invariant behaviour often associated with circular groove cavities. These experimentally demonstrated properties directly support application directions in reflective structural colour, compact colour filtering, frequency-selective reflective surfaces, and optical-variable anti-counterfeiting features.
\end{abstract}

\maketitle

\section{Introduction}

Plasmonic metasurfaces make it possible to tailor reflection, absorption, and colour response at subwavelength scales by structuring metallic surfaces into resonant optical meta-atoms~\cite{anker2008biosensing,lindquist2012engineering,meinzer2014plasmonic, yu2014flat, kildishev2013planar}. In the visible and near-infrared, such surfaces can display optical characteristics that differ strongly from bulk metals, including selective absorption, engineered reflection minima, and geometry-defined structural colour~\cite{zhang2011continuous,cheng2015structural,keshavarz2017review,kristensen2016plasmonic,daqiqeh2020nanophotonic,clausen2014plasmonic}. Among the most attractive platforms are continuous-metal plasmonic cavities, since they can combine compactness, fabrication simplicity, and robust optical response without relying on disconnected antenna elements or multilayer dielectric coatings~\cite{ding2018review, genevet2017recent}.\\

V-shaped nanogrooves are a particularly important class of plasmonic cavities because they support gap-surface-plasmon (GSP) modes with strong field confinement near the groove bottom and sidewalls~\cite{bozhevolnyi2005channel,bozhevolnyi2006effective,smith2015gap,ding2018review}. When such grooves are bent into closed or nearly closed cavity geometries, they can sustain more complex localized resonances related to circulating plasmon waves and cavity-mode interference~\cite{vesseur2009modal,vesseur2011plasmonic}. In periodic arrays, these cavity resonances can generate narrow or broad reflection features that are tunable by groove geometry and can be exploited in colour filtering, frequency-selective optical surfaces, and related nanophotonic devices~\cite{zhang2011continuous,clausen2014plasmonic,cheng2015structural}.

Most reported groove-cavity systems have focused on geometries with high in-plane symmetry, especially circular resonators. Circular cavities are valuable because they can show a stable response and only weak azimuthal dependence under normal illumination~\cite{vesseur2009modal,vesseur2011plasmonic}. However, that same symmetry can limit the ability to encode orientation-dependent optical information. A natural extension is therefore to introduce symmetry breaking while preserving the continuous-metal cavity topology. Trifolium nanocavities offer such a route: their three-lobed profile retains a compact groove-like resonator geometry but introduces preferred in-plane directions, distinct local curvatures, and a central coupling region where different cavity sectors can hybridize.

Another important aspect is the plasmonic metal platform itself. Single-crystal Au(111) provides a smoother and lower-defect optical surface than conventional polycrystalline films, thereby reducing extrinsic damping and improving resonance reproducibility~\cite{Radha2012,Kuttge2008,Mejard2017,Schmidt2012,Greenwood2022}. In related Au(111)-based groove metasurfaces, the optical response has already been shown to depend strongly on groove depth and to support application-relevant spectral selectivity~\cite{Kuttge2009,smith2015gap,zhang2011continuous}. Here we build on that single-crystal platform and present a compact experimental study of trifolium nanocavity metasurfaces. The purpose is not to model the structures numerically, but to establish the experimentally observed optical behaviour and identify application directions that follow directly from the measured spectra. In particular, we show that the trifolium metasurfaces exhibit strong depth-tunable reflection response and a measurable azimuth-dependent spectral variation, making them promising for reflective structural-colour~\cite{cheng2015structural, keshavarz2017review, kristensen2016plasmonic, daqiqeh2020nanophotonic, kumar2012printing} and optical-variable surface functionalities~\cite{smith2017metal, vu2025optical, Jung2021}.

\section{Theory}
The optical response of the trifolium nanocavities can be understood from the physics of gap-surface-plasmon ($gsp$) modes supported by tapered metallic grooves. For a moderately wide groove with $w > (\lambda \varepsilon_d)/(\pi |\varepsilon_m|)$ such that $\left|\frac{-2\varepsilon_d}{w \varepsilon_m}\right| < k_0$, the dispersion relation of $gsp$ modes $k_{gsp}$ supported by V-grooves, can be approximated as~\cite{bozhevolnyi2008scaling}
\begin{equation}
k_{gsp} \approx k_0 \sqrt{\varepsilon_d + \frac{2\varepsilon_d \sqrt{\varepsilon_d - \varepsilon_m}}{k_0 w (-\varepsilon_m)}},
\label{eq:kgsp_linear}
\end{equation}
where $k_0=2\pi/\lambda$ is the free-space wavenumber, and $\varepsilon_d$ and $\varepsilon_m$ are the dielectric and metal permittivities, respectively. Equation~\eqref{eq:kgsp_linear} shows that the effective mode index $\eta_{eff} = k_{gsp}/k_0$, determining mode confinement and dispersion, increases as the groove narrows, so that the deepest and narrowest parts of the cavity support the strongest confinement and field localization~\cite{bozhevolnyi2006effective,smith2015gap}. 

In the present structures, these confined groove modes are folded into a compact three-lobed cavity rather than a straight channel. The measured reflection resonances are therefore interpreted as hybridized cavity modes sustained by coupling between the three groove sectors through the central junction~\cite{smith2015gap}. Increasing the cavity depth enhances confinement and increases the effective optical path length, which naturally accounts for the experimentally observed redshift and reshaping of the reflection minima. This picture is closely related to the whispering-gallery-like behaviour reported for closed groove cavities, where plasmon waves circulate along the cavity contour and form geometry-dependent resonances~\cite{vesseur2009modal,vesseur2011plasmonic}. In the trifolium geometry, however, the continuous rotational symmetry of the circular cavity is broken by the three-lobed profile. The resulting structure introduces preferred in-plane directions and modifies the overlap between the incident polarization and the cavity current distribution, thereby producing the observed azimuth-dependent response.

\section{Materials and methods}
\subsection{Fabrication}
 Single-crystalline Au(111) microplates were synthesized via air-thermolysis using gold(III) chloride trihydrate (Sigma-Aldrich) and tetraoctylammonium bromide (ToABr, Alfa Aesar) as precursors, without further purification~\cite{Radha2012}. The process involved phase transfer of \ch{[AuCl_4]^-} ions from an aqueous 25~mM gold precursor solution (0.059~g in 5~mL Milli-Q water) to an organic phase using 50~mM ToABr in toluene (0.137~g in 5~mL). The solutions were mixed in a 1:2 volume ratio (e.g., 450~$\mu$L aqueous: 900~$\mu$L organic), stirred for phase transfer, and the resulting orangish-red organic phase was drop-cast onto cleaned glass substrates. Thermal treatment at 130~\degree C for 5~days yielded hexagonal and triangular gold microplates after 2--3~days. The resulting microplates are optically thick, beyond the skin depth (typically $\sim 25-30$ nm in the visible), on the scale relevant to visible plasmonic reflection measurements and therefore act as effectively opaque gold templates. Trifolium nanocavity arrays were then patterned into the top Au surface by focused Ga$^+$ ion beam milling using a Helios NanoLab 600 (FEI) operated at \SI{30}{kV} and \SI{28}{pA}. Focused Ion Beam (FIB) milling is a high-precision nanofabrication technique used to create V-groove cavities by sputtering material from a substrate~\cite{smith2015gap,langford2007focused,lindquist2012engineering}. Representative microscopy of the fabricated metasurface in Fig.~\ref{fig:fabrication} confirms the three-lobed cavity morphology.

Each trifolium nanocavity comprises three elongated groove lobes joined at a common central junction, thereby forming a cavity with threefold rotational symmetry. The V-groove top width is \SI{100}{nm}, with lobe lengths in the range of \SIrange{0.56}{0.60}{\micro\meter} measured from the central junction to the tip, and maximum lobe widths in the range of \SIrange{0.28}{0.30}{\micro\meter}, as determined by scanning electron microscopy (SEM). The trifolium unit cell had a lateral footprint of approximately $0.9\times 1.2\,\mu$m, and the center-to-center pitch is approximately \SI{1.35}{\micro\meter}, with the fabricated square arrays containing $15\times15$ cavities. Two cavity depths, \SI{300}{nm} and \SI{350}{nm}, were investigated. 

\subsection{Optical characterization}
Optical characterization was performed in reflection mode using linearly polarized broadband white-light illumination from a halogen source (Ocean Optics HL-2000-FHSA) in a microscope-based K\"ohler geometry. The reflected signal was collected with a Zeiss Achrostigmat objective (20$\times$, NA = 0.4) and analysed using an Avantes AvaSpec SensLine spectrometer coupled to a WITec SNOM microscope. Owing to the finite numerical aperture, the measurements correspond to near-normal incidence. Spectra were recorded with an exposure time of 10~ms, averaged over ten acquisitions, dark-spectrum subtracted, and normalized to a protected silver-mirror reference spectrum; reflection from unstructured flat Au(111) was also recorded for comparison. The measured spectral range covered approximately \SIrange{450}{900}{nm}. To assess the role of cavity orientation, the sample was rotated in-plane and reflection spectra were recorded for selected azimuthal angles. In this paper, the analysis is restricted to the measured reflectance response and its dependence on azimuthal orientation and groove depth.\\
\begin{figure}[t]
    \centering
    \includegraphics[width=0.95\linewidth]{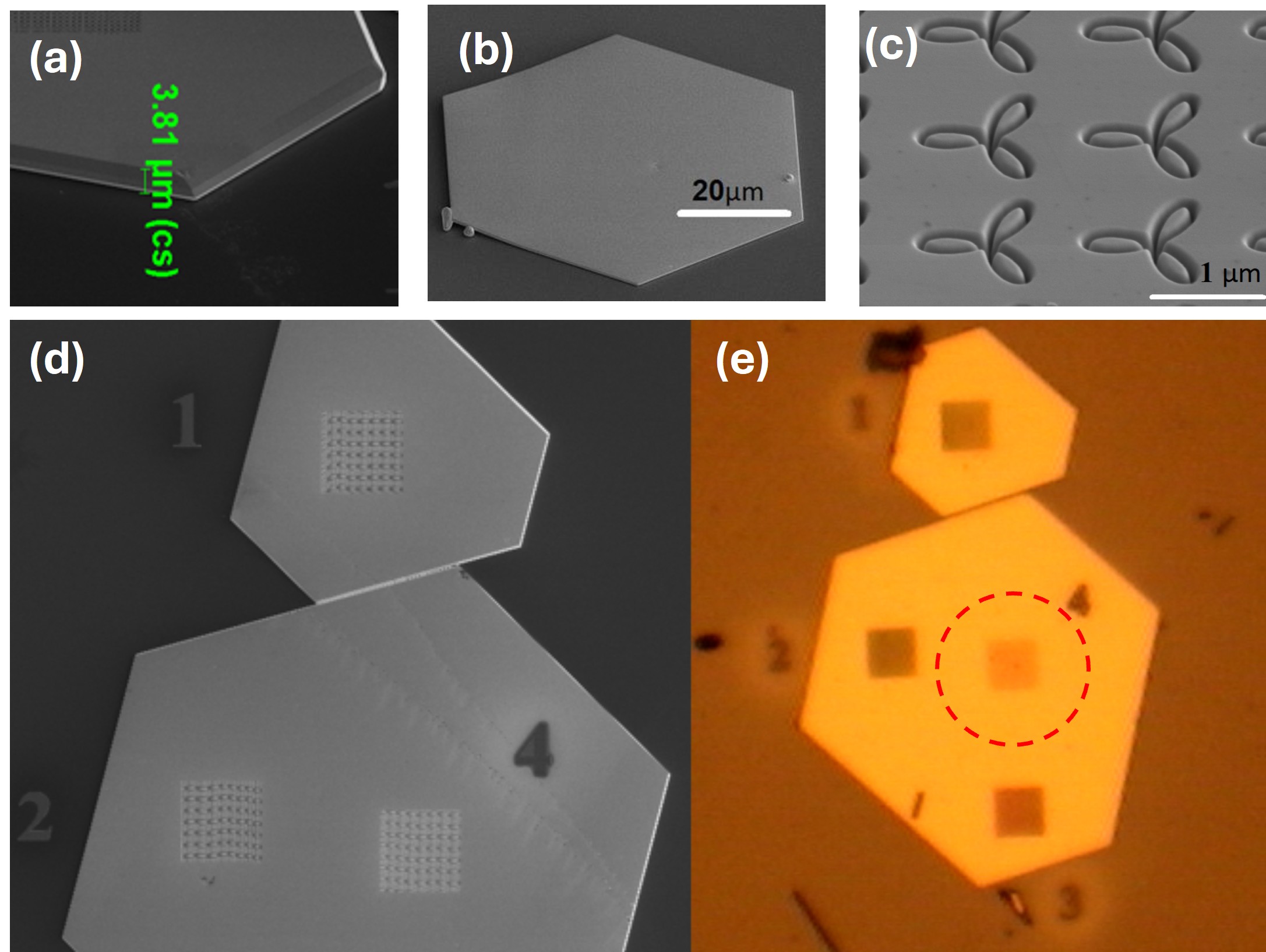}
    \caption{Structural characterization of trifolium nanocavity metasurfaces fabricated on single-crystal Au(111) microplates. (a) Cross-sectional image of a representative Au microplate with a thickness of approximately \SI{3.81}{\micro\meter}. (b) SEM image of a hexagonal Au microplate before patterning. (c) Magnified SEM image of the periodic trifolium nanocavity array; scale bar: \SI{1}{\micro\meter}. (d) SEM overview of several patterned microplates containing square arrays of trifolium cavities. (e) Optical micrograph of the patterned plates, where the dashed circle marks the structured area.}
    \label{fig:fabrication}
\end{figure}

Figure~\ref{fig:fabrication} summarizes the fabrication procedure and subsequent morphological characterization of the trifolium nanocavity metasurfaces. The single-crystalline Au(111) microplates employed as plasmonic substrates exhibit lateral dimensions on the order of several tens of micrometres and a thickness of approximately \SI{3.81}{\micro\meter}, as shown in Fig.~\ref{fig:fabrication}(a,b). FIB milling is used to define periodic arrays of trifolium-shaped cavities with high fidelity, as confirmed by the higher-magnification SEM image in Fig.~\ref{fig:fabrication}(c). The overview image in Fig.~\ref{fig:fabrication}(d) illustrates that multiple patterned arrays can be reproducibly fabricated across different microplates, while the optical micrograph in Fig.~\ref{fig:fabrication}(e) indicates that the structured regions are easily visible under optical inspection due to their altered reflectivity. When the metasurface is illuminated with focused white light (red dashed circle in Fig.~\ref{fig:fabrication}(e)), the illuminated region displays a distinct colour compared to both the surrounding nearby structured regions and the unstructured Au(111) background. The images presented in Fig.~\ref{fig:fabrication} demonstrate the high structural quality of the Au microplates and validate the reliable realization of the trifolium cavity geometry over well-defined array regions.

\section{Results and discussion}
Across the entire visible–near-infrared spectral range, the structured surfaces deviate markedly from the reference Au response, confirming that the trifolium cavities introduce pronounced plasmonic spectral selectivity. In both structured cases, the spectra exhibit a low-reflectance region at shorter wavelengths, a broad high-reflectance response in the red spectral region ($\lambda = 600-700$ nm), and a distinct long-wavelength reflection minimum band in the near-infrared, $\lambda = 700-800$ nm.

\subsection{Azimuth-dependent optical response}
\begin{figure}[t]
    \centering
    \includegraphics[width=0.98\linewidth]{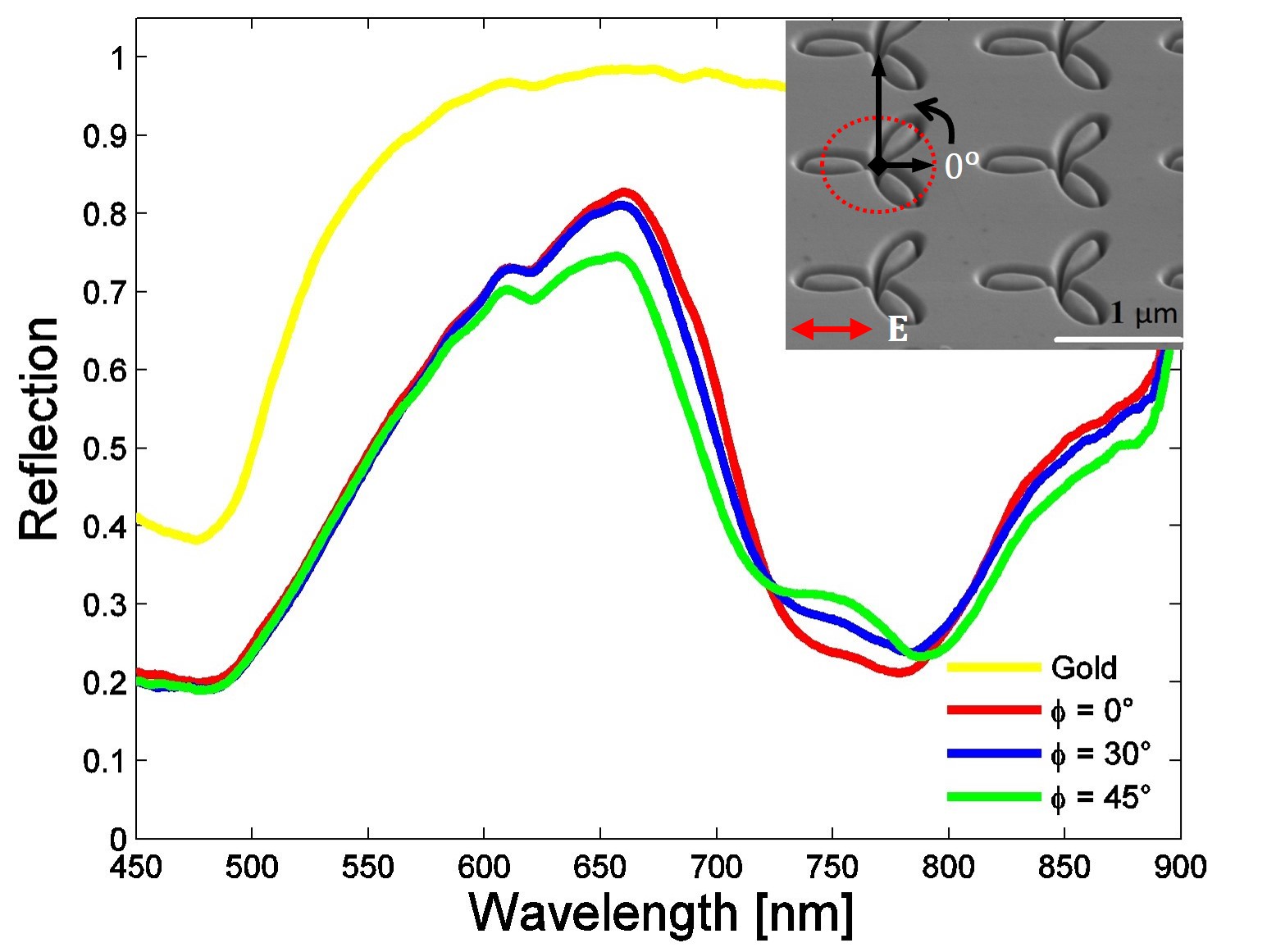}
  \caption{Measured normalized reflection spectra of flat Au and trifolium nanocavity metasurfaces for azimuthal orientations $\phi=0^\circ$, $30^\circ$, and $45^\circ$. Compared with unstructured gold, the patterned surfaces display a broad reflection band in the red spectral region and a pronounced minimum at longer wavelengths. The modest but clear evolution of the spectra with $\phi$, especially in the \SIrange{730}{800}{nm} range, reveals an orientation-dependent plasmonic response arising from the reduced in-plane symmetry of the trifolium cavity. The inset shows an SEM image of the array and the definition of the rotation angle relative to the fixed electric-field $E$ direction.}
    \label{fig:azimuth}
\end{figure}
\begin{figure}[t!]
    \centering
    \includegraphics[width=0.98\linewidth]{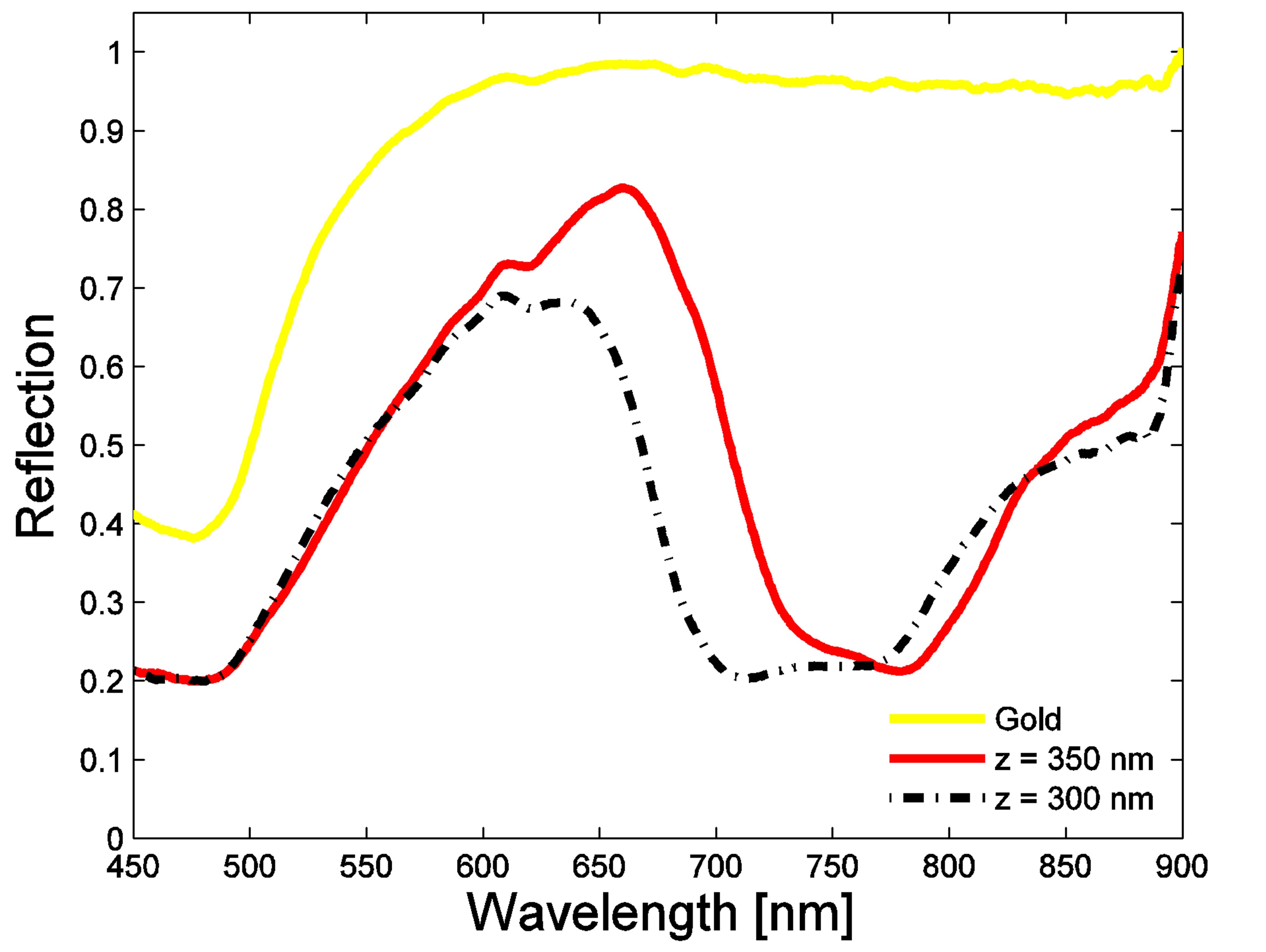}
  \caption{Measured reflection spectra of unstructured single-crystal Au(111) and trifolium nanocavity metasurfaces with cavity depths of \SI{300}{nm} and \SI{350}{nm} for $\phi = 0\degree$. The structured surfaces exhibit a substantially altered optical response relative to unstructured gold. An increase in cavity depth produces a pronounced redshift of the long-wavelength reflection minimum band, $\lambda = 700-800$ nm, and significantly modifies the intermediate spectral region, $\lambda = 600-680$ nm, demonstrating a strongly depth-dependent spectral response.}
    \label{fig:depth}
\end{figure}
Figure~\ref{fig:azimuth} demonstrates that the trifolium nanocavity metasurface exhibits a distinct, though moderate, dependence of its optical response on the in-plane azimuthal orientation of the sample. The spectral features comprise a broad, high-reflectance band centered approximately within the \SIrange{600}{680}{nm} region, as well as a pronounced reflection minimum at longer wavelengths in the \SIrange{730}{800}{nm} range. Upon rotating the sample from $\phi = 0^\circ$ to $30^\circ$ and $45^\circ$, the overall spectral lineshape remains essentially unchanged, indicating that the same underlying set of plasmonic cavity modes is predominantly excited. Nonetheless, the depth, width, and precise spectral resonance position of the long-wavelength minimum band exhibit a clear azimuthal dependence.

This behaviour is a direct optical consequence of the reduced in-plane symmetry of the trifolium cavity. Unlike a circular groove resonator, which is nearly rotationally symmetric and therefore often only weakly sensitive to azimuth under normal illumination~\cite{vesseur2009modal,vesseur2011plasmonic}, the trifolium shape introduces preferred orientations for coupling between the incident electric field and the cavity current distribution. The three-lobed geometry likely redistributes plasmonic excitation among different sectors of the cavity as the sample is rotated, producing the modest but reproducible spectral variation observed experimentally.

\subsection{Depth-dependent reflection response}
Figure~\ref{fig:depth} presents the normalized reflection spectra of the unstructured single-crystal Au(111) together with those of trifolium metasurfaces milled to depths of \SI{300}{nm} and \SI{350}{nm}. The most salient experimental observation is the pronounced dependence on groove depth. Increasing the cavity depth from \SI{300}{nm} to \SI{350}{nm} induces a clear redshift of the dominant long-wavelength minimum and substantially modifies the intermediate spectral band. This modification cannot be described as a simple rigid translation of the spectrum. Instead, both the relative prominence of the resonance mode around \SIrange{580}{680}{nm} and the depth of the long-wavelength minimum band evolve with the cavity depth. This behaviour is characteristic of plasmonic groove resonators, in which the effective mode index (from Eq.~\eqref{eq:kgsp_linear}), modal path length, and the degree of electromagnetic confinement are determined by the cavity geometry~\cite{bozhevolnyi2006effective,smith2015gap,Kuttge2009,ding2018review}.

\subsection{Application implications of the measured spectra}
The spectra in Fig.~\ref{fig:azimuth} show only a modest spectral displacement of the long-wavelength minimum band, on the order of \SIrange{10}{11}{nm}, while more clearly modifying the minimum reflectance and overall lineshape as the azimuth angle changes. By contrast, Fig.~\ref{fig:depth} shows that the long-wavelength reflection minimum redshifts from approximately \SI{716}{nm} for the \SI{300}{nm}-deep cavities to about \SI{779}{nm} for the \SI{350}{nm}-deep cavities, corresponding to a net shift of roughly \SI{63}{nm}. These results indicate that cavity depth is the primary control parameter for resonance position, whereas in-plane rotation mainly affects coupling strength and spectral contrast.\\

These optical responses point to three immediate application directions. First, the measured response is relatively broadband rather than ultranarrow, which is advantageous for {compact reflective colour filters} and related {frequency-selective reflective surfaces}~\cite{zhang2011continuous,clausen2014plasmonic,daqiqeh2020nanophotonic}. Broad, geometry-controlled reflection minima and maxima are often more useful for practical appearance control than extremely sharp resonances, since they provide robust spectral tailoring and stronger modification of the intrinsic gold background. The trifolium cavities therefore act as efficient absorptive and/or scattering resonators that redistribute reflected intensity across the visible and near-infrared, a behaviour that is widely exploited in plasmonic colour surfaces and reflective optical filters~\cite{cheng2015structural,keshavarz2017review,kristensen2016plasmonic,daqiqeh2020nanophotonic}.\\

Second, and most distinctively, the azimuth-dependent response demonstrates that the trifolium metasurface carries not only a static reflection signature but also {orientation-dependent optical information}. The controlled spectral change observed under in-plane rotation is particularly attractive for {optical-variable security features}, including anti-counterfeiting tags, reflective authentication elements, and other surface-integrated nanophotonic security devices. Anti-counterfeiting tags~\cite{smith2017metal, vu2025optical} are specialized, secure identifiers (e.g., smart barcodes, optical physical unclonable function (PUF), holograms, etc) applied to products to verify authenticity, prevent counterfeiting, and enable supply chain traceability. In such applications, the combination of wavelength-selective reflection and orientation encoding is highly desirable because it produces a visual or spectroscopic signature that is difficult to replicate without reproducing the underlying nanoscale geometry~\cite{Jung2021}.

Third, the pronounced depth dependence of the reflectance spectra shows that the spectral behaviour of the trifolium metasurface can be tuned purely through geometry. This is a key requirement for {reflective structural-colour engineering}, where the visible appearance is controlled by nanostructure design rather than by pigments, dyes, or multilayer interference coatings~\cite{cheng2015structural,keshavarz2017review,kristensen2016plasmonic,daqiqeh2020nanophotonic, kumar2012printing}. In the present case, varying the cavity depth shifts and reshapes the dominant reflection features over a broad spectral range, indicating that the reflected colour and overall optical appearance can be engineered in a predictable and fabrication-compatible manner.

\section{Conclusion}
We have experimentally demonstrated plasmonic metasurfaces composed of trifolium-shaped nanocavities milled into single-crystal Au(111) microplates and investigated their reflection response in the visible--near-infrared. The measured spectra show that the optical response is strongly tunable by groove depth, with the deeper cavities exhibiting a redshifted long-wavelength minimum and a reshaped intermediate spectral band. In addition, the trifolium geometry produces a measurable azimuth-dependent spectral variation under in-plane rotation, revealing the practical optical consequence of symmetry breaking in a continuous-metal cavity architecture.

These findings define the main contribution of the present paper. The trifolium metasurface combines geometry-controlled reflective spectral selectivity with optical-variable behaviour in a compact and fabrication-friendly single-crystal gold platform. Based on the measured optical properties alone, the most immediate application directions are reflective structural colour, compact colour filtering, frequency-selective reflective surfaces, and anti-counterfeiting optical-variable devices. More broadly, the work shows that moving beyond circular cavity symmetry offers a simple experimental route to enrich the functionality of groove-based plasmonic metasurfaces.

\begin{acknowledgments}
The author acknowledges the support, facilities, and technical assistance associated with the synthesis of Au(111) microplates, focused-ion-beam nanofabrication, and optical reflection measurements. In particular, the author thanks DAAD for funding this project, M.\ R.\ Goncalves for technical support in the laboratory, and G.\ Neusser for support during the FIB fabrication and SEM imaging of the metasurfaces. The study was done at Ulm University, Germany, during the author’s MSc studies under the supervision of Prof.\ O.\ Marti, supported by the Tanzania--Germany Postgraduate Training Program.
\end{acknowledgments}
\subsection*{Conflicts of Interest}
The author declares no conflicts of interest.
\subsection*{Data Availability Statement}
Data supporting the results presented in this paper are not publicly available at this time but may be obtained from the corresponding author upon reasonable request.

\bibliographystyle{apsrev4-2}
\bibliography{myreferences_trifolium}

\end{document}